\documentclass[fleqn,usenatbib]{mnras}

\usepackage{newtxtext,newtxmath}
 
\usepackage[T1]{fontenc}

\DeclareRobustCommand{\VAN}[3]{#2}
\let\VANthebibliography\thebibliography
\def\thebibliography{\DeclareRobustCommand{\VAN}[3]{##3}\VANthebibliography}

\usepackage{graphicx}	
\usepackage{amsmath}	
\usepackage{multirow}   
\usepackage{booktabs}

\title[JWST/MIRI source count model]{Cosmic star-formation history and black hole accretion history inferred from the \textit{JWST} mid-infrared source counts}

\author[S. J. Kim et al.]{
Seong Jin Kim,$^{1}$\thanks{E-mail: seongini@gmail.com}
Tomotsugu Goto,$^{1,2}$
Chih-Teng Ling,$^{1}$
Cossas K.-W. Wu,$^{1,2}$
Tetsuya Hashimoto,$^{3}$
\newauthor
Ece Kilerci,$^{4}$
Simon C.-C. Ho,$^{5,6,7,8}$ 
Yuri Uno,$^{3}$
Po-Ya Wang,$^{1,2}$
Yu-Wei Lin$^{1,2}$
\\
$^{1}$Institute of Astronomy, National Tsing Hua University, 101, Section 2. Kuang-Fu Road, Hsinchu, 30013, Taiwan (R.O.C.)\\
$^{2}$Department of Physics, National Tsing Hua University, 101, Section 2. Kuang-Fu Road, Hsinchu, 30013, Taiwan (R.O.C.)\\
$^{3}$Department of Physics, National Chung Hsing University, 145, Xingda Road, Taichung, 40227, Taiwan (R.O.C.)\\
$^{4}$Sabanc{\i} University, Faculty of Engineering and Natural Sciences, 34956, Istanbul, Turkey\\
$^{5}$Research School of Astronomy and Astrophysics, The Australian National University, Canberra, ACT 2611, Australia\\
$^{6}$Centre for Astrophysics and Supercomputing, Swinburne University of Technology, P.O. Box 218, Hawthorn, VIC 3122, Australia\\
$^{7}$OzGrav: The Australian Research Council Centre of Excellence for Gravitational Wave Discovery, Hawthorn, VIC 3122, Australia\\
$^{8}$ASTRO3D: The Australian Research Council Centre of Excellence for All-sky Astrophysics in 3D, ACT 2611, Australia}

\date{Accepted 2023 November 7. Received 2023 November 7; in original form 2022 December 15}

\pubyear{2015}
 
\begin{document}
\label{firstpage}
\pagerange{\pageref{firstpage}--\pageref{lastpage}}
\maketitle

\begin{abstract}
With the advent of the James Webb Space Telescope (\textit{JWST}), extra-galactic source count studies were conducted down to sub-$\mu$Jy in the mid-infrared (MIR), which is several tens of times fainter than what the previous-generation infrared (IR) telescopes achieved in the MIR.  In this work, we aim to interpret the \textit{JWST} source counts and constrain cosmic star-formation history (CSFH) and black hole accretion history (BHAH).
We employ the backward evolution of local luminosity functions (LLFs) of galaxies to reproduce the observed source counts from sub-$\mu$Jy to a few tens of mJy in the MIR bands of the \textit{JWST}. 
The shapes of the LLFs at the MIR bands are determined using the model templates of the spectral energy distributions (SEDs) for five representative galaxy types (star-forming galaxies, starbursts, composite, AGN type 2 and 1).  By simultaneously fitting our model to all the source counts in the six MIR bands, along with the previous results, we determine the best-fit evolutions of MIR LFs for each of the five galaxy types, and subsequently estimate the CSFH and BHAH. 
Thanks to the \textit{JWST}, our estimates are based on several tens of times fainter MIR sources, the existence of which was merely an extrapolation in previous studies.   
    
\end{abstract}
 

  

\section{Introduction}

Extragalactic source counts (defined as the number count of sources as a function of  flux density) have been used as one of the basic cosmological tests because it is simple and straightforward. It is a practical tool to check the variation of the source distribution with increasing luminosity distance (or redshift). 
In the early studies in the infrared (IR), such as  \citet{Elbaz1999}, \citet{Serjeant2000}, and \citet{Pearson2005}, the detection limit at 15$\mu$m was at the sub-milliJansky (sub mJy) level based on the 
\textit{Infrared Space Observatory} (ISO) observations in the 1990s.
Thanks to the AKARI \citep{Murakami2007}  and Spitzer \citep{Werner2004} space telescopes, the detection limit became deeper, reaching a few tens of $\mu$Jy in subsequent studies by  \citet{Pearson2010}, \citet{Takagi2012}, and \citet{Pearson2014}.

Much deeper IR science data from the James Webb Space Telescope \citep[\textit{JWST},][which was launched in late December 2021]{Gardner2006, Kalirai2018}, has been publicly available, e.g., Stephan's Quintet\footnote[1]{\url{https://webbtelescope.org/contents/media/images/2022/034/01G7DA5ADA2WDSK1JJPQ0PTG4A}}  and the  Cosmic Evolution Early Release Science Survey\footnote[2]{\url{https://ceers.github.io/overview.html}} (CEERS; Finkelstein et al. 2017)  since July 2022. 
This availability allowed for new source count (SC) tests using very faint ($S_{\nu}<\mu$Jy)  mid-infrared (MIR) sources.
Ling et al. (2022, 2023) presented the source counts  at the F770W, F1000W and F1500W bands of the Mid-IR Instruments \citep [MIRI,][]{Rieke2015} using early data from the Stephan's Quintet field. Additionally, Wu et al. (2023) utilized data from the CEERS field to provide additional source counts at the F1280W, F1800W, and F2100W bands.  In the previous SC studies with the former IR telescopes \citep[e.g.,][]{Elbaz1999, Serjeant2000, Malkan2001, Lagache2005, Pearson2005, Pearson2010}, one of the key issues was understanding the properties of the sources contributing to the cosmic infrared (IR) background \citep[CIB,][]{Puget1996, Lagache2005}. 
The recent \textit{JWST} source counts have reached the flux limits  below the $\mu$Jy level (e.g.,  at 7.7, 10, and 12.8 $\mu$m; Ling et al. 2022, 2023; Wu et al. 2023), revealing galaxies several tens of times fainter.  
The \textit{JWST} achieves 0.3 arcsec resolution at 7.7$\mu$m (F770W), while the Spitzer has a point spread function (PSF) of 2 arcsecs at 8 $\mu$m (IRAC 4) \citep{Gardner2006, Werner2004}. Therefore, the \textit{JWST} has seven times better resolution compared to the Spitzer space telescope. The sources contributing to the IR background in previous works are now resolved into discrete sources.

Half, at least, of the energy in the Universe, is emitted in the IR wavelengths, with a peak in the far-IR, providing insight into the overall bulk of the IR energy \citep[i.e., total IR luminosity,][]{Lutz2014, Casey2014}.
In the mid-IR (MIR) regime, where emission is dominated by rather warmer/hotter dust heated by strong radiation, various emission features exist, such as polycyclic aromatic hydrocarbons (PAHs) or silicate features, which offer clues to the detailed dust properties associated with star-formation (SF) activity \citep[][Ohyama et al. 2018]{Tielens2008, Desai2007, Takagi2010,  Kim2019, Sirocky2008}. Additionally, MIR observations probe the hot dust heated by active galactic nucleus (AGN) activity \citep{Gruppioni2008, Wang2020, Lacy2021}. Importantly, the advantage of IR observation is that the data is less affected by dust obscuration.

Luminosity function (LF) is a statistical description of the distribution of galaxy luminosities and is defined as the number of galaxies in a specific volume as a function of luminosity (Johnston 2011). It is a measure of the number density of galaxies at different luminosities, and provides important information about the populations of galaxies and their evolution \cite[e.g.,][]{Elbaz1999, R-Robinson2009, Goto2010,  Gruppioni2011}.  One of the key features of the luminosity function is its shape, which is affected by a variety of evolutionary factors such as star formation activity, galaxy evolution and growth, the role of supermassive black holes (SMBHs), and so on \citep[e.g,][Gruppioni et al. 2013]{Lagache2003}.  
More accurate measurement of LFs has been one of the main concerns in observational cosmology, to obtain more accurate cosmic star-formation history and galaxy evolution \citep[e.g.,][Kim et al. 2015; Kilerci Eser \& Goto 2018; Goto et al. 2019]{Gruppioni2010,  Goto2015}. 
The source count (SC) is closely related to the luminosity function of galaxies, which describes the distribution of galaxies in terms of observed brightness \citep[e.g.,][]{Gruppioni2011, Pearson2014}. If redshift information is obtained (measured or assumed), astronomers can construct multi-wavelength luminosity functions by combining information from source counts at different wavelengths. This can provide a more complete picture of the galaxy populations. Therefore, the source count is an important tool in studying galaxy evolution and luminosity functions.

Previous source count studies have shown that IR luminosity functions have to change dramatically with redshift and the "no-evolution model" can not describe the bump feature around 0.1-1mJy in the source counts \citep{Elbaz1999, Serjeant2000, Lagache2003, Pearson2010}.  Given the crucial role of evolution, modelling IR source counts becomes essential to understand the impact of galaxy evolution on counts and gain insights into the dusty star formation history \citep{R-Robinson2009, Gruppioni2011}.  
In this work, we take much fainter galaxies into account to understand the connection between galaxy source counts and cosmic evolution history.
Our primary goal is to reproduce the observed source counts across the contiguous MIRI bands of the \textit{JWST}, and derive constraints on the cosmic star formation history (CSFH) and black hole accretion history (BHAH). We aim to develop a model that effectively describes the faint end ($\sim$ sub-$\mu$Jy) of the MIR source counts obtained by the \textit{JWST}, while also considering previous results from the former IR space telescopes that cover up to a few tens of mJy.  Utilizing the model we obtain, we interpret source count results in terms of the evolution of different galaxy populations contributing to the source counts across a wide range of flux densities. Additionally, the CSFH and BHAH will be updated using new parameters obtained including very faint IR galaxies. 

This paper is structured as follows. In Section \ref{sec2}, we summarise the source counts from the \textit{JWST} as well as from previous works. Section \ref{sec3} shows how to derive our source count model in the MIR and compares them with observed source counts. In section \ref{sec4}, we discuss the model we obtain and infer CSFH and BHAH. We summarise our work in section \ref{sec5}, 
In the model computation,  we assume the Salpeter initial mass function (IMF, Salpeter 1955) and $\Lambda$ cold dark matter ($\Lambda$CDM) cosmology with H$_0$ = 70km s$^{-1}$  Mpc$^{-1}$, $\Omega_{\rm m} = 0.3$, and $\Omega_{\Lambda}$ = 0.7.

\section{Observed MID-IR source counts}
\label{sec2}

\subsection{Source counts from the \textit{JWST}/MIRI}

The early release science (ERS) data from the \textit{JWST}/MIRI have given us brand new source counts for faint ($S_{\nu} < \mu$Jy) mid-IR sources  (Ling et al. 2022, 2023; Wu et al. 2023) at the six broad band filters (F770W, F1000W, F1280W, F1500W, F1800W, and F2100W) of the \textit{JWST}/MIRI. 
Ling et al. (2022, 2023) used the two sub-fields (i.e., jw02732-o002\_t001 and jw02732-o006\_t001, covering  $\sim$1.9 arcmin$^2$ and $\sim$2.7 arcmin$^2$, respectively) of the Stefan's Quintet observations, by masking out the five main galaxies,  to detect much fainter extragalactic sources in the fields, at F770W, F1000W, and F1500W.
Wu et al. (2023) used the CEERS field data and took the two sub-fields (i.e., jw01345-o001 and jw01345-o002) covering $\sim$2 arcmin$^2$ and $\sim$4 arcmin$^2$  observed by six MIRI filters.
They used \textsc{Source-Extractor} V2.19.5 \citep{Bertin1996} to extract sources. Background estimation for photometry was carried out based on the `\textsc{Photutils}'\footnote[1]{\url{https://photutils.readthedocs.io/en/stable/}} package. 
In their works, the source count results were given in the flux range brighter than 80 \% completeness limit in each band (e.g., about  0.8  and 2.5 $\mu$Jy at F1000W and F1500W, respectively).

\begin{figure}
  \includegraphics[width=\columnwidth]{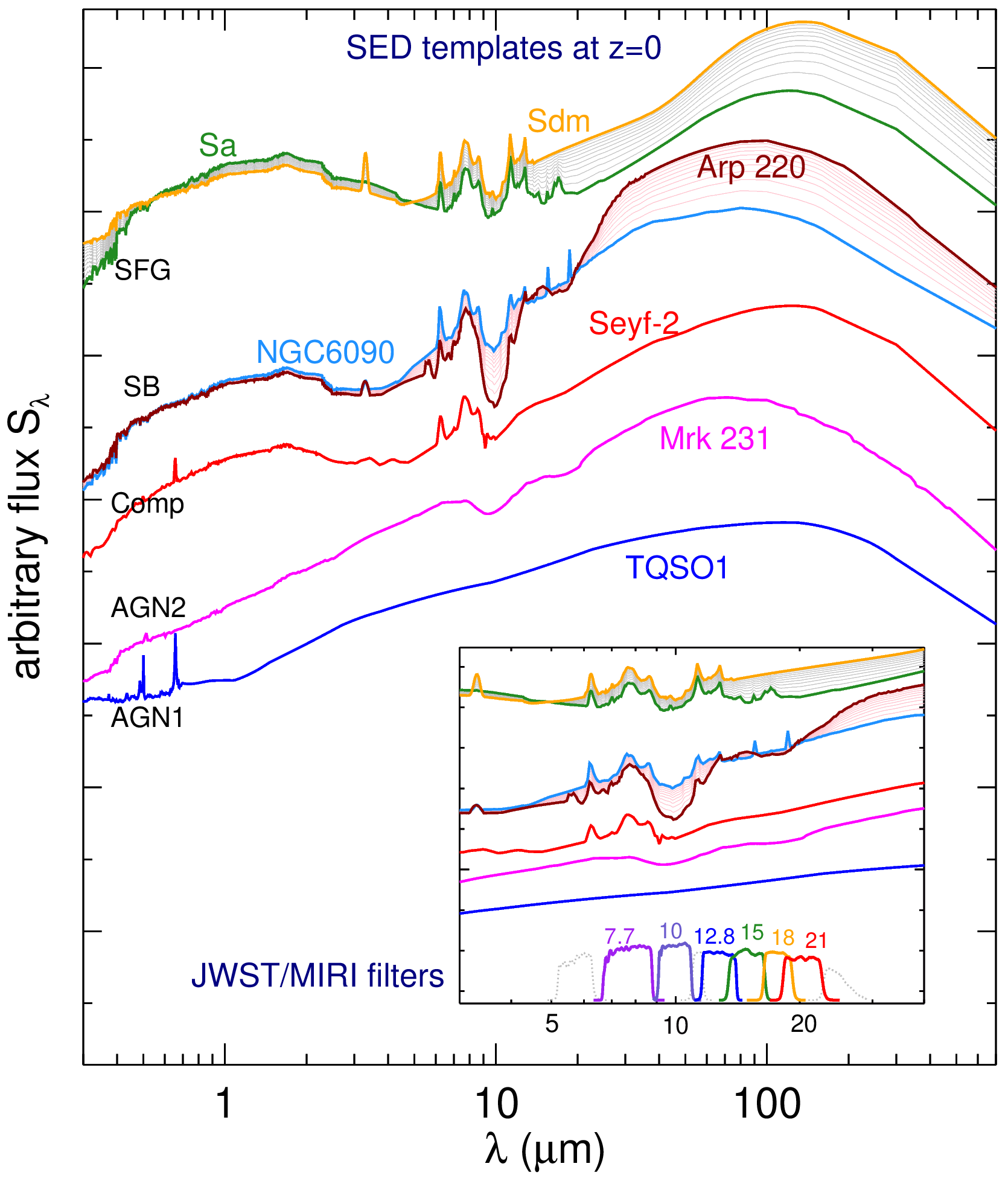}
   \caption{Model spectral energy distribution (SED) templates at $z=0$ for five representative galaxy populations from \citet{Polletta2007}.
    Green/orange: normal star-forming galaxy (SFG) population represented by Sa and Sdm with intermediate SEDs (shown in grey) between them. Intermediate SEDs are generated by the combination of the two templates.  Cyan/dark red: starburst (SB) population represented by NGC 6090 and Arp 220 with intermediate SEDs (shown in pink). Red: composite galaxy population represented by Seyfert 2 -- a mixture of SFG and low luminosity AGN (LLAGN).  Magenta: AGN type-2 (obscured), and blue: AGN type-1 (unobscured).  In a small box, we display the MIR range (3-40$\mu$m) along with the \textit{JWST}/MIRI filter bands.
    }
    \label{f1_SEDtempl}
\end{figure}

\begin{figure*}
  \includegraphics[width=\textwidth]{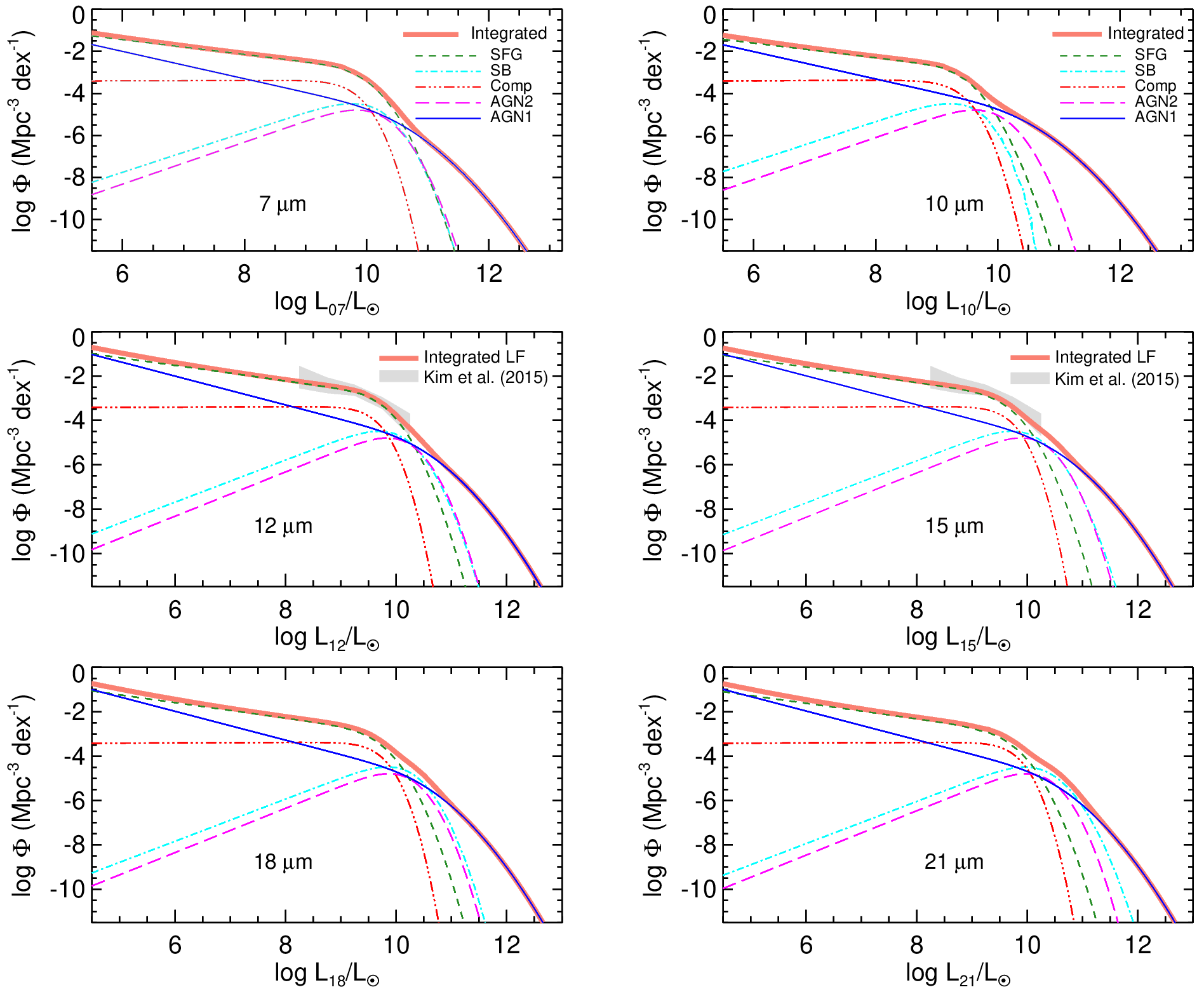}
    \caption{Model local luminosity functions (LLFs) in the six mid-IR (MIR) bands of the \textit{JWST}/MIRI, for five representative galaxy types, which are derived based on the parameters given for 15$\mu$m LLF \citep[Table 1 of ][]{Gruppioni2011}. Green: normal star-forming galaxy (SFG) given by spiral, cyan: starburst (SB), red: composite (a mixture of SFG and low luminosity AGN), magenta: type-2 AGN, and blue: type-1 AGN. The thick salmon line indicates the integrated LF of these five types. For 12$\mu$m and 15 $\mu$m bands, we compare the model LFs with observed LFs (grey regions) based on AKARI's NEP survey data (Kim et al. 2015).   } 
    \label{f2_LLFs}
\end{figure*}

Their source count results showed roughly reasonable agreement with the extension/prediction of the previous models, which tells us that our understanding of the IR Universe has been consistent with previous expectations.   However, there are slight deviations in their comparisons, especially in the fainter range, which a simple extension of the old models can not give sufficient explanations for.
A large error bar of source count in a certain flux bin comes from a small number of sources contributing to the flux bin, which is determined based on the Poisson error for small-number statistics \citep{Gehrels1986}. Their observed number counts show occasional scatter and fluctuation of the data points in a certain range of flux (Wu et al. 2023), however, it is not certain if it is because of statistical variance/fluctuation between the small patches of the different sky fields (cosmic variance).  
We attempt to see how the evolutionary parameters have to be changed/adjusted to describe the new source count results obtained from the \textit{JWST} observations.

\subsection{Source counts from the previous IR telescopes}

Since Oliver et al. (1997)  provided the 15$\mu$m ISOCAM source counts from the HDF field, 
many MIR source count studies have been carried out based on the ISOCAM \citep[][etc.]{Elbaz1999, Gruppioni2002, Pearson2005}. AKARI's large area survey \citep[e.g., North Ecliptic Pole,][]{Wada2008, Kim2012}  and Spitzer \citep[e.g., EGS, SWIRE, and CDF-S field, ][]{Fazio2004, Shupe2008, Papovich2004} data were used to derive the MIR source counts \citep{Takagi2010, Pearson2010, Pearson2014, Murata2014, Davidge2017}.    
In this work, we also use the MIR source counts from previous works when we derive new evolution parameters.  The previous source counts from the AKARI bands provide good comparison samples because wavelength coverage is similar to that of MIRI, and some of them have almost the same effective wavelengths (e.g., 15 and 18 $\mu$m).  In this work, we accept the previous source counts data only where the completeness for the source detection is higher than 80\%.

\begin{figure*}
\includegraphics[width=\textwidth]{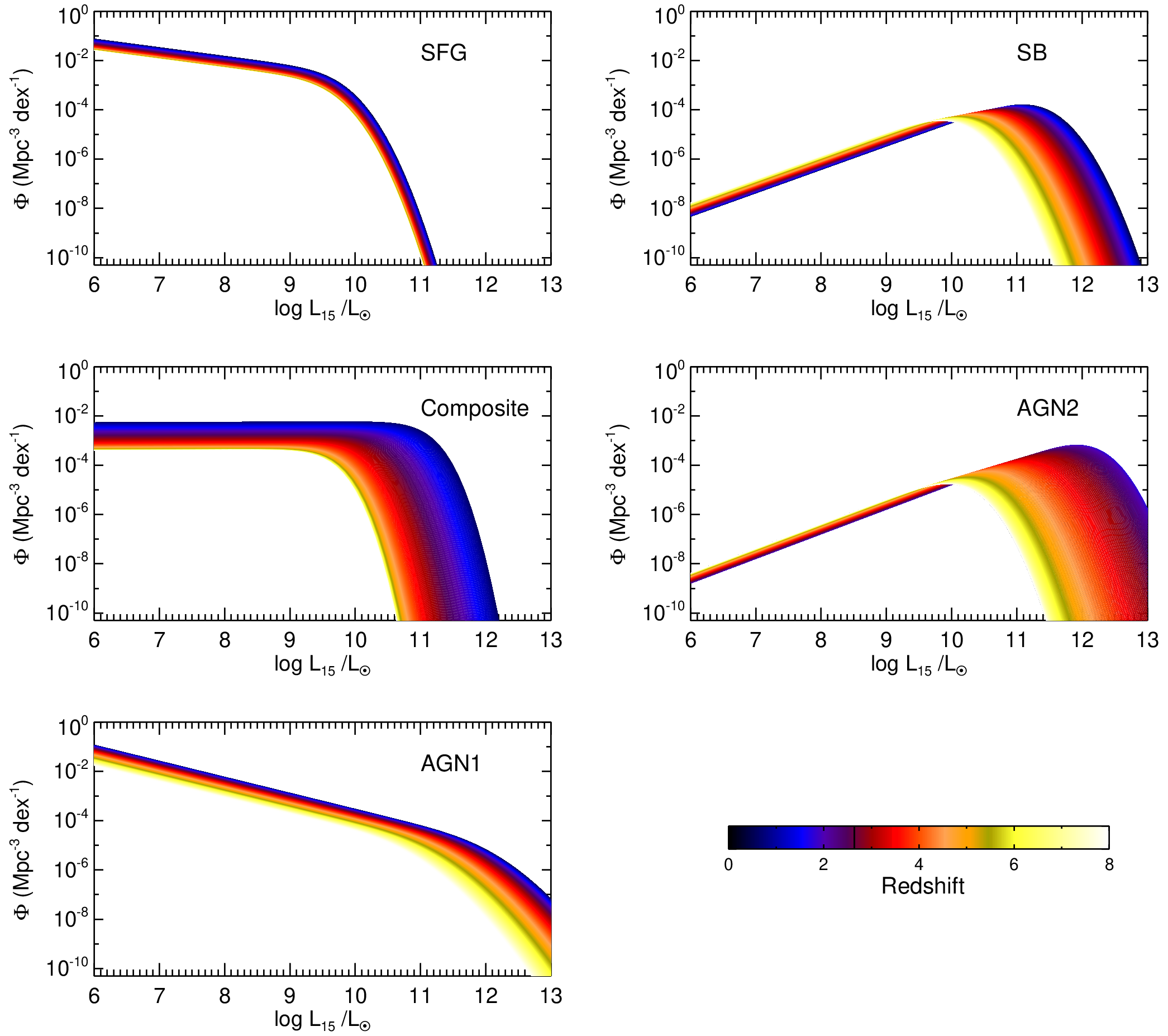}
 \caption{The evolution of the local luminosity functions (LLFs) at 15$\mu$m band (F1500W), according to equations (2) and (3), with redshift represented by a colour gradient from dark blue through red to yellow (refer to the colour bar on the bottom right).  
 Luminosity bin size is 0.1 dex, and the redshift interval is 0.02, in our computation. Normal star-forming galaxies (SFGs, top-left panel), starbursts (SB, top-right panel), composite (a mixture of SFG and LLAGN, middle-left panel), AGN type-2 (middle-right panel) and AGN type-1 (bottom-left panel) are shown. }
    \label{f3_evLLFs}
\end{figure*}

\begin{figure*}
	\includegraphics[width=\textwidth]{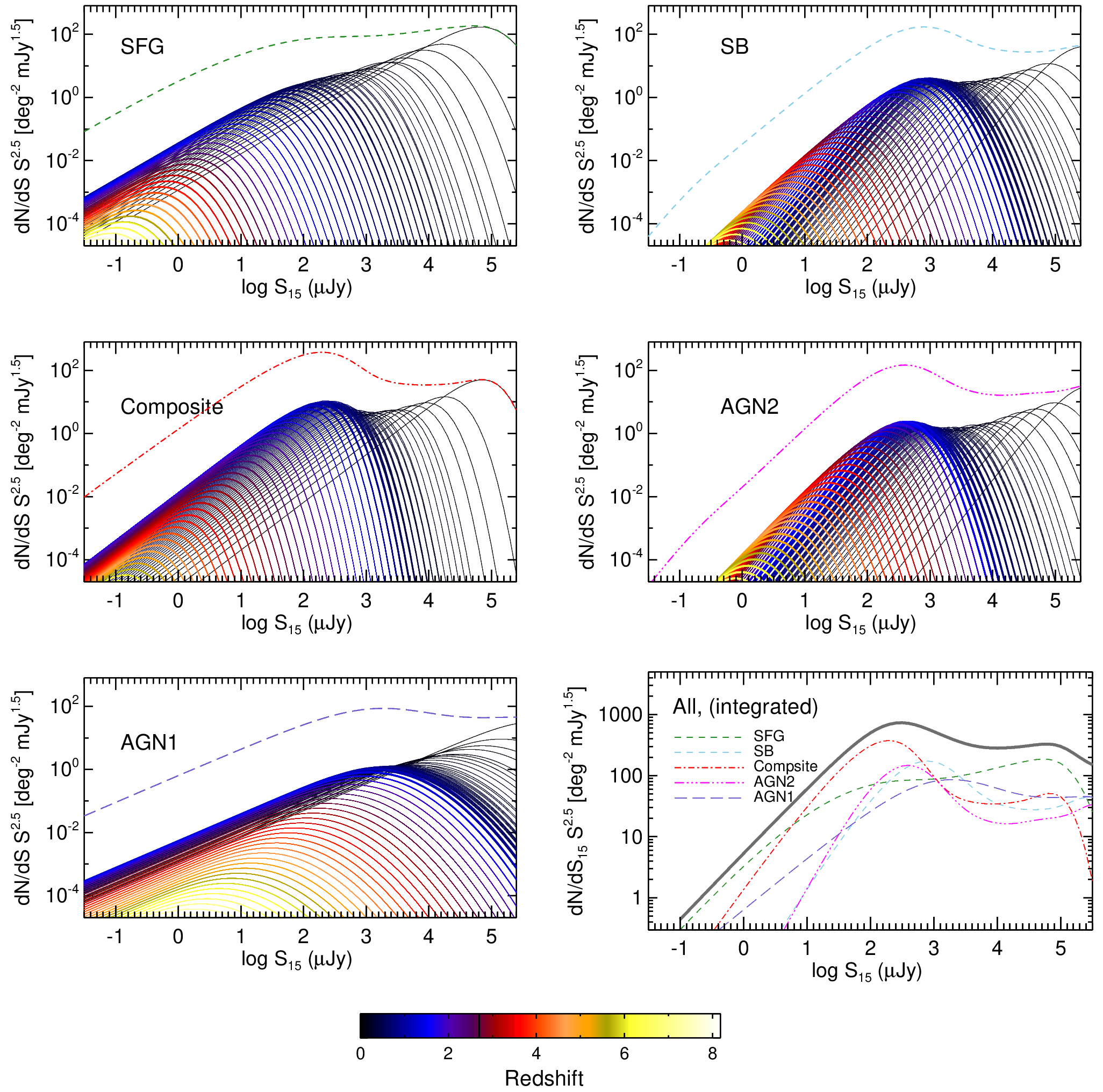}
    \caption{The evolution of number counts. Here, the x-axis is given in unit of flux density ($\mu$Jy) at 15 $\mu$m, which is converted from luminosity (the x-axis in Fig. \ref{f3_evLLFs}) based on the K-correction using the 15 $\mu$m filter (F1500W).  Each panel shows the evolution of each galaxy type, with redshift represented by a colour bar on the bottom.  The broken line in each panel shows integrated values over all redshifts --  SFG (top-left panel), starbursts (SB,  top-right panel), composite (a mixture of SFG and LLAGN, middle-left panel), AGN type-2 (middle-right panel), and AGN type-1 (bottom left). The bottom right panel shows the sum of all the types. }.
    \label{f4_evEucl_SC}
\end{figure*}

\begin{figure*}
  \includegraphics[width=\textwidth]{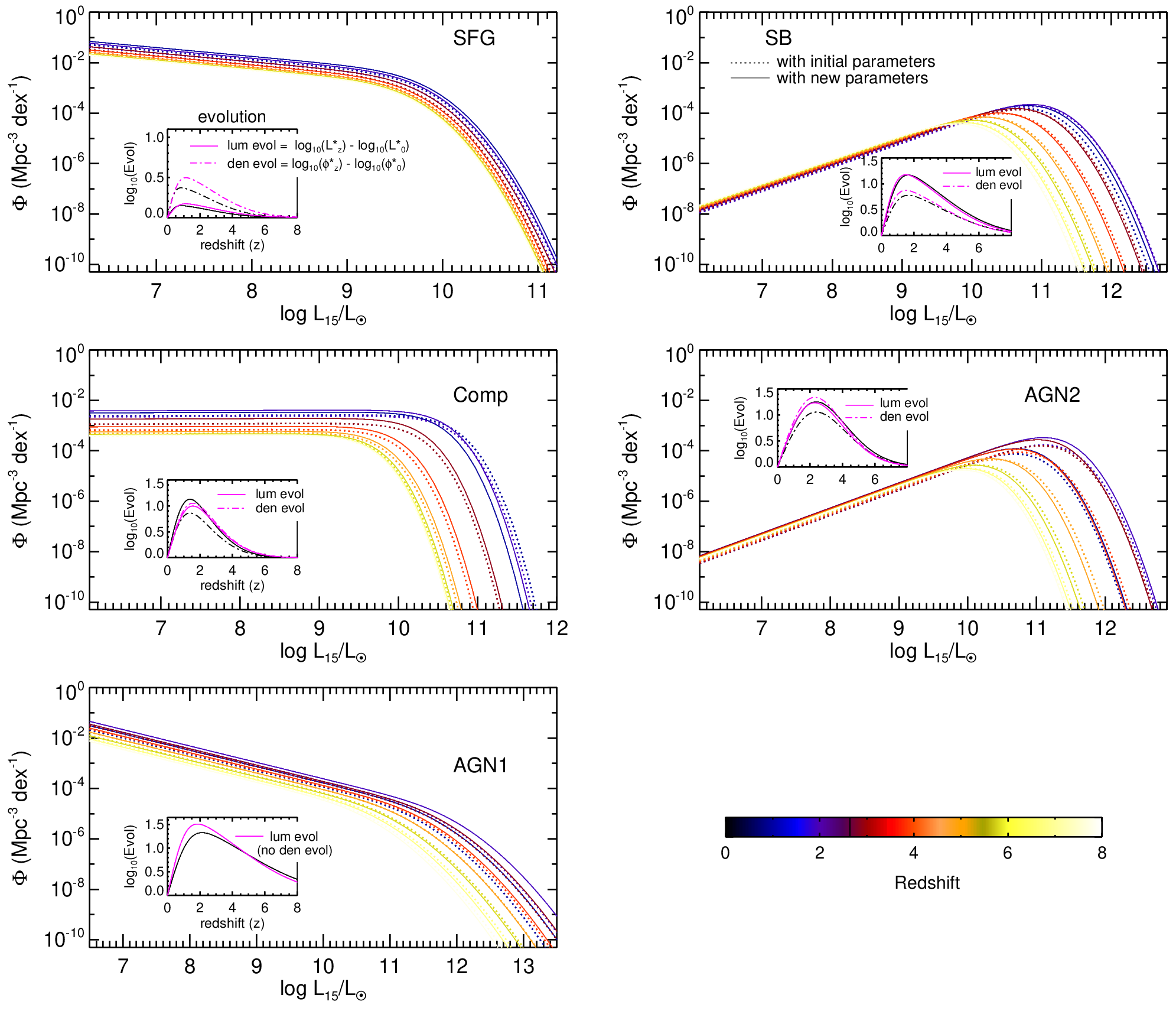}
    \caption{Comparison of LLF evolutions between the initial parameters (shown in Fig. \ref{f3_evLLFs}) and new parameters obtained from the fitting in this work. We show 15$\mu$m LFs at redshifts $z$ = 1.0, 2.0, 3.0, ..., 7.0, and 8.0 only, for better visibility. The dotted curves indicate the LFs with initial parameters, which are from Figure \ref{f3_evLLFs}. The solid curves are the LFs with newly derived parameters. The inset (a small box) in each panel shows luminosity evolution (solid curve) and density evolution (dot-dashed curve) as a function of redshift. The y-axis `$\rm{log}_{10}(Evol)$' represents the evolution terms as a function of $z$, except for the first terms ($\rm{log}_{10}L^\star (0)$ and  $\rm{log}_{10} \Phi^\star (0)$) in equations 2 and 3. Therefore, luminosity evolution is $\rm{log}_{10}L^\star (z)$-$ \rm{log}_{10}L^\star (0)$. In this small box, black curves denote evolution with initial parameters, and magenta curves show evolution with new parameters. Therefore, LFs in dotted curves are based on the evolution in black, while LFs in solid curves are based on the magenta curves in the insets.
    }
    \label{f5_evLLFs2}
\end{figure*}

\section{Source count model}
\label{sec3}

\subsection{Local luminosity functions and evolution models}

To obtain physical motivations on the parameters of LFs, we attempt to model/reproduce  the mid-IR source counts by combining the
\textit{JWST}/MIRI source counts (Ling et al. 2022, 2023; Wu et al. 2023) with those obtained by previous works \citep[Oliver et al. 1997;][]{ Serjeant2000, Pearson2010, Takagi2012, Pearson2014, Davidge2017}.
 
We use one of the simple/popular ways, taking the backward evolution method by \citet {Gruppioni2011}, where evolution models are described by evolution parameters given for different galaxy types.
The evolution of the local luminosity functions (LLFs) towards high redshifts has been one of the frequently used methods \citep[e.g.,][etc.]{R-Robinson2009, Bethermin2011, Gruppioni2011} because this empirical approach reasonably fits the observed galaxy number counts and is consistent with theories.  
\citet{Gruppioni2011} present the model LLFs  for mid-IR bands (e.g., 12 and 15 $\mu$m): ${\Phi}(L)$, 
 described by the following function \citep[][modified Schechter function]{Saunders1990}:
\begin{equation}
\Phi(L)=\Phi^{\star}\left(\frac{L}{L^{\star}}\right)^{1-\alpha} \mathrm{exp}\left[-\frac{1}{2\sigma^2} \mathrm{log}_{10}^2\left(1+\frac{L}{L^{\star}}\right)\right]
\end{equation}
which behaves as a power law for $L<<L^{\star}$ and as a Gaussian in $\mathrm{log}L$ for $L>>L^{\star}$.  One of the main reasons we use the methods from \citet{Gruppioni2011} is that we can take advantage of testing different types of galaxies (see Fig. \ref{f1_SEDtempl}) because \citet[Fig. 1 and Table 1 of][]{Gruppioni2011} provide the LLFs for five representative populations: normal star-forming galaxies (SFGs), starbursts (SB), composites -- a mixture of SFG and low-luminosity AGN (LLAGN),  AGN type-2 (obscured AGN; AGN2), and AGN type-1 (unobscured AGN; AGN1). 

\begin{figure*}
  \includegraphics[width=\textwidth]{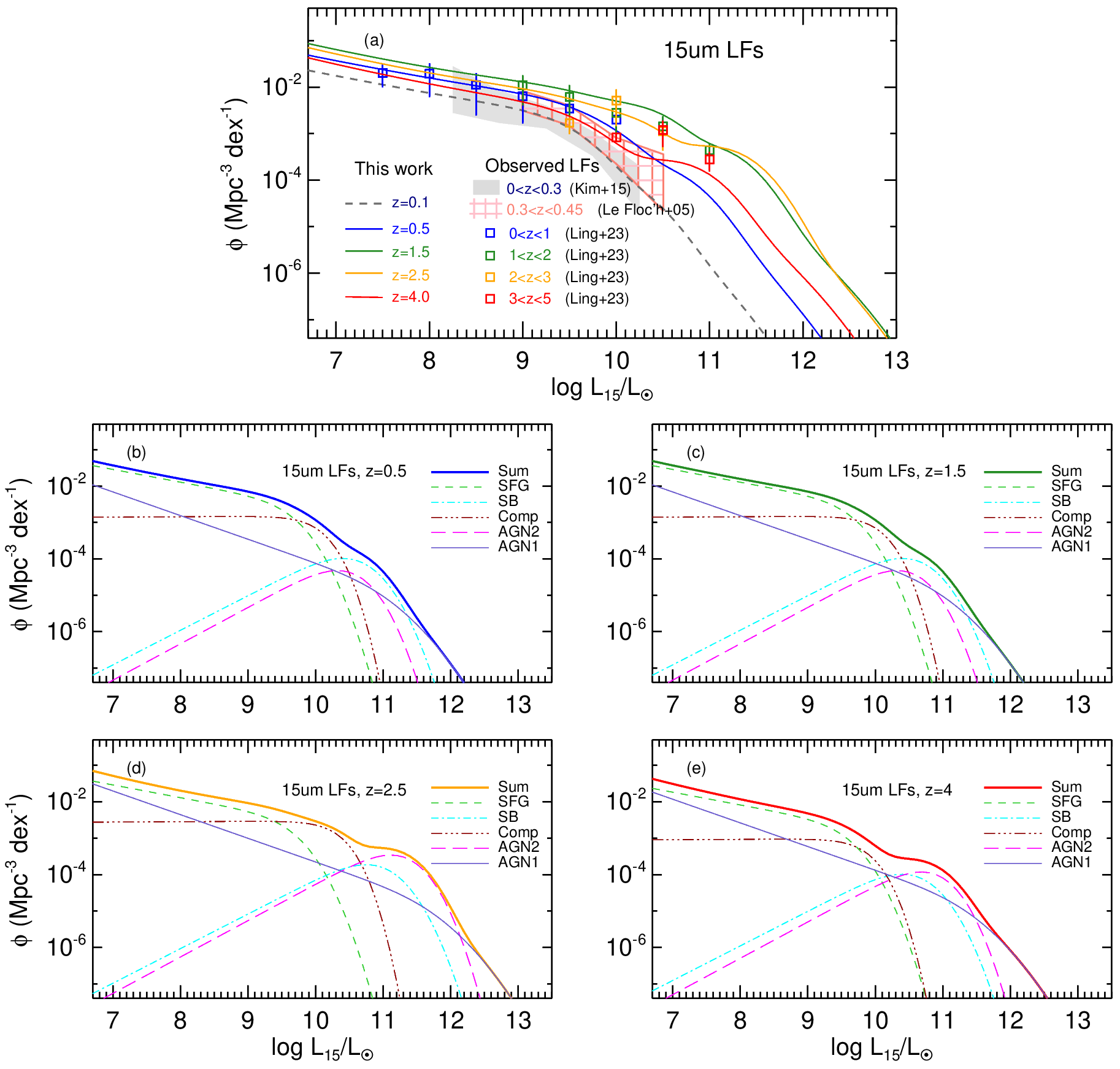}
    \caption{Panel (a): Comparison of 15$\mu$m LFs - our model LFs at different redshifts (curves), against the observed 15$\mu$m LFs. Our model LFs are based on the new evolution parameters obtained through the fitting to the SCs.     
    A grey shaded area indicates observed LLF from Kim et al. (2015, AKARI/NEP-Wide field), and the meshed area is from Le Floc'h et al. (2005, Spitzer/CDFS field), and open squares from Ling et al. 2023 (submitted, JWST/CEERS field).   Blue, green, orange and red lines indicate the LF at redshift $z=0.5, 1.5, 2.5$, and $4.0$, respectively. They are the sum of the LFs for each population at each redshift -- see panels (b), (c), (d), and (e), which show how the LFs on the top panel are generated. Single LFs for each population in lower panels   are a subset of LFs explained in Fig.\ref{f5_evLLFs2}.    } 
    \label{f6_LF15obs}
\end{figure*}

\begin{figure*}
  \includegraphics[width=\textwidth]{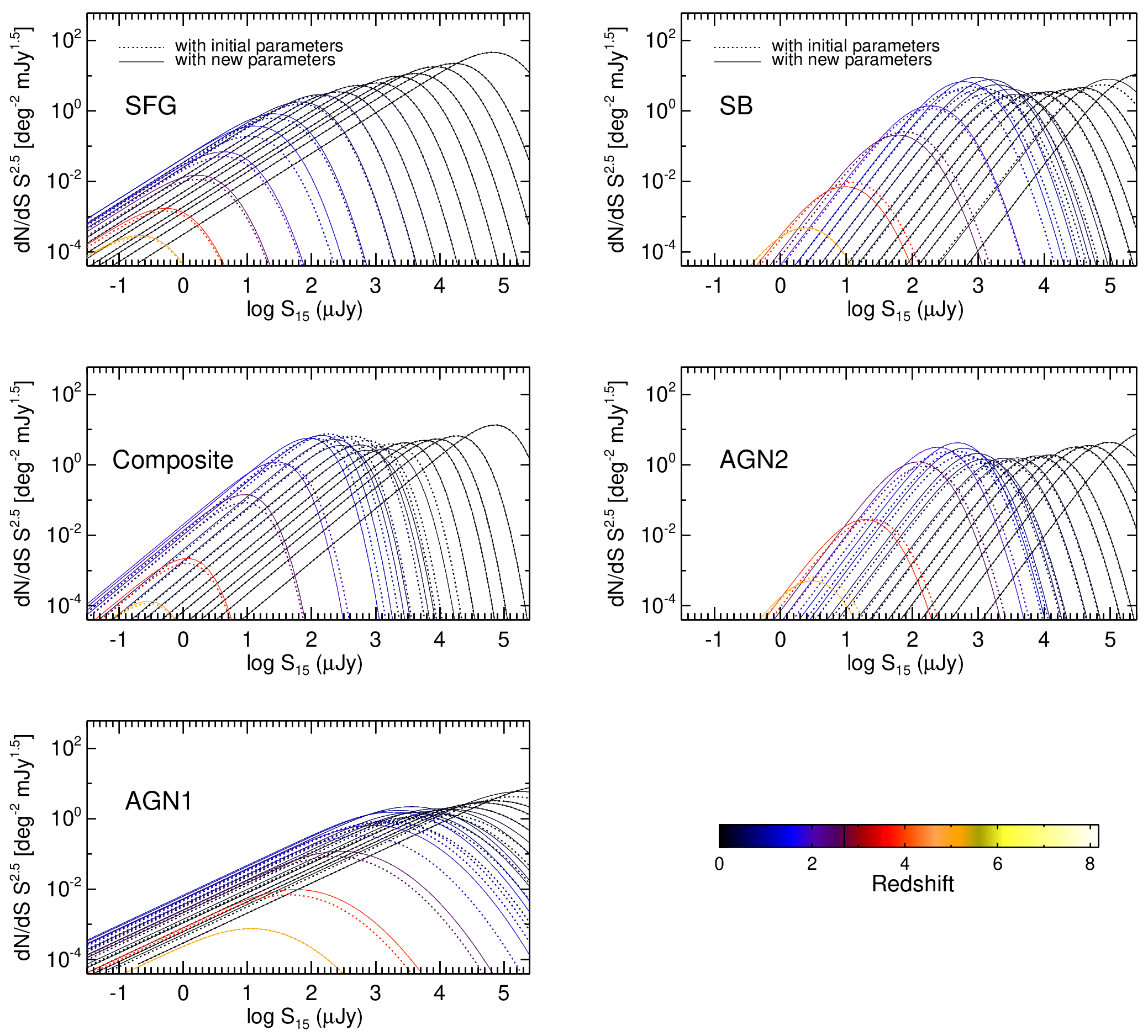}
    \caption{Comparison of the 15 $\mu$m galaxy number count models at different redshifts between the initial parameters (as given in Fig. \ref{f4_evEucl_SC}) and the new parameters obtained from the fitting in this work. The dotted curves indicate the number counts models with initial parameters (as given in Fig. \ref{f4_evEucl_SC}). Solid curves are number counts based on the new parameters. The solid curves represent the source count based on the newly obtained parameters. These curves show slight differences due to the changes in parameters. 
      } 
    \label{f7_evNCs2a}
\end{figure*}

 First, we obtain LLFs for six MIR bands of the \textit{JWST}/MIRI (F770W, F1000W, F1280W, F1500W, F1800W, and F2100W) -- based on the given information concerning 15$\mu$m LLFs for each galaxy population \citep[i.e., the LF parameters such as $L^\star$, $\Phi^\star$, $\alpha$, and $\sigma$ in Table 1 of][]{Gruppioni2011},  the LLFs for the other wavelength bands are derived using the filter bands of the MIRI and model spectral energy distributions (SEDs). We can reproduce the model 15$\mu$m LLFs presented in \citet{Gruppioni2011}  using the filter response of the F1500W and galaxy model SEDs \citep[e.g.,][]{Polletta2007}. While doing this, we can also use the other MIRI filters (F770W, F1000W, F1280W, F1800W, and F2100W) to obtain LLFs at these bands at the same time.

For this procedure, we take the SED model templates from \citet[][]{Polletta2007}, as shown in Fig. \ref{f1_SEDtempl}. 
As in \citet{Gruppioni2011},  we applied the template variations for SFG and SB according to the luminosity range, where the slope of the luminosity function changes rapidly. For the SFG population, we take the assumption that an SED evolves with luminosity from Sa ($L_{\rm {15\mu m}} < 10^{9} \rm L_{\odot}$) to Sdm  ($L_{\rm {15\mu m}} \ge 10^{10} \rm L_{\odot}$).  For the SB population, the SED  varies from a moderate NGC6090 ($L_{\rm {15\mu m}} \le 10^{10} \rm L_{\odot}$) up to extreme Arp 220 ($L_{\rm {15\mu m}} > 10^{11.9} \rm L_{\odot}$).   The template variation is discrete as shown in Fig. 1.  
For the composites, AGN1, and AGN2 populations, the templates of Seyfert-2, TQSO1, and Mrk231 are taken, respectively.   
Using these templates, we obtained local luminosity functions (LLFs).
In Fig. \ref{f2_LLFs}, we show the model LLFs at the six JWST/MIRI bands (7.7, 10, 12.8, 15, 18, and 21$\mu$m) for five different galaxy types.

\subsection{Evolution of LLFs and number counts}

For the evolution of the LLFs for galaxy populations, we follow the evolution model with the parameterization that regulates the shapes of the evolution functions as shown in \citet{Gruppioni2011}: 
\begin{eqnarray}
\mathrm{log}_{10}(L^\star(z))=\mathrm{log}_{10}(L^\star(0)) + \frac{A_L}{\omega \sqrt{2\pi}} e^{-z^2/2\omega^2}\times \mathrm{erf}\left(\kappa \frac{z}{\omega}\right) \label{equation:evoll} \\
\mathrm{log}_{10}(\Phi^\star(z))=\mathrm{log}_{10}(\Phi^\star(0)) + \frac{A_{\Phi}}{\omega \sqrt{2\pi}} e^{-z^2/2\omega^2}\times \mathrm{erf}\left(\kappa \frac{z}{\omega}\right) \label{equation:evolr}
\end{eqnarray} 
 where $\mathrm{erf}(x)=\frac{2}{\sqrt{\pi}}\int^{x}_{0}e^{-t^2}\mathrm{d}t$ is the ``error function''. $A_L$ and $A_{\Phi}$ are the normalisations for luminosity evolution $L^\star(z)$ and density evolution $\Phi^\star(z)$, respectively.  $\kappa$ regulates the shape of the function, whose combinations with  $\omega$ (scale factor) define the evolution peak and skewness (asymmetric shape) of the function towards high redshift.

In this work, the model LLFs derived in the six bands of the \textit{JWST} evolve in terms of luminosity and density up to $z\sim$8 according to equations (2) and (3).
\citet[][in their Table 2]{Gruppioni2011} present these 4 parameters ($A_L$, $A_{\Phi}$, $\kappa$, and $\omega$) for 5 different types of galaxies, i.e., 20 parameters in total.
Here, we aim to update/renew these evolution parameters by fitting the faint end of the MIR source counts obtained from the \textit{JWST}. We attempt to encompass a broader redshift range, reaching up to $z$=8 (which corresponds to the reionisation era) in our calculation, while \citet{Gruppioni2011}  covered up to $z$=5. 

Therefore, in this work, the mixture of the five types of galaxies is given in terms of their LFs \citep{Gruppioni2011}. 
They evolve differently according to 
equations (2) and (3) with 20 parameters. For example, we show the evolution of the 15 $\mu$m LLFs for five galaxy types (SFG, SB, composite, and AGN2 and AGN1) in Fig. \ref{f3_evLLFs}. 
The redshift interval is 0.02, and the flux density interval is given in 0.1 dex in the calculation.

If we convert the luminosity to the observed flux density, then each panel will give the number of galaxies as a function of flux density ($\rm S_{\nu}$) in the observed frame at 15 $\mu$m as shown in Fig. \ref{f4_evEucl_SC} ($\Delta\log S_{\nu}=0.1$). 
The sum of all the individual SC  is given in a broken line on the top of each panel: green dashed (SFG), cyan dashed (starburst), red dot-dashed (composite), magenta dot-dot-dashed (AGN2) and blue dashed (AGN1).  These plots clearly show that the brightest galaxies tend to be local ones (given in black/blue) whereas the fainter galaxies tend to be at higher redshift (in red/orange). Especially for SB and AGN2, high redshift galaxies are dominant in number at the faint end.
On the bottom right panel, these five summed LFs for each galaxy type are finally integrated into a thick grey line.  This grey line shows the distribution of all galaxies which are accumulated and projected to the observed frame. Therefore, counting the number of galaxies along this grey curve is the same as counting the number of galaxies projected on the night sky. These number counts of the model galaxies are shown in Euclidean normalised values ($\rm dN/dS \,S^{2.5}$). We will compare these SC models  with the observed source counts in each band (in Sec 4.1).

\begin{table*}
\centering
 \caption{Parameter changes after the fitting procedures - the evolutionary parameters as the initial guess in this work (the best estimates for the parameters in \citet{Gruppioni2011}) have changed as shown in the table. Fitting results/errors are from a software package, `\textsc{MPFIT}'.  Most of them are fitted within the parameter ranges we set, except for $A_{\phi}$ for SB which reached the imposed limit (8.50) to prevent catastrophic behaviour. Gr+11 denotes \citet{Gruppioni2011}.  }
 \label{table1}
\begin{tabular}{|c|c|c|c|c|c|c|c|c|}
\hline
\multirow{2}{1em}{ } & \multicolumn{2}{|c|}{\textbf{$A_L$}} & \multicolumn{2}{|c|}{\textbf{$A_{\phi}$}} & \multicolumn{2}{|c|}{\textbf{$\omega$}} & \multicolumn{2}{|c|}{\textbf{$\kappa$}}   \\
\cmidrule(lr){2-3} \cmidrule(lr){4-5} \cmidrule(lr){6-7} \cmidrule(lr){8-9}
   & Gr+11 & This work & Gr+11 & This work & Gr+11 & This work & Gr+11 & This work \\
\hline
SFG  & 1.0 & 1.01$\pm$0.02 & 2.5 & 3.79 $\pm$0.37 &2.5 & 2.40$\pm$0.31 &5.3 & 5.11$\pm$0.02  \\
\hline
Starburst  &12.0 & 11.95$\pm$0.02 &8.0 & 8.50 (limit) & 3.5 & 3.50$\pm$0.02 &3.0 & 3.10$\pm$0.02   \\
\hline
Composite\ &8.5 & 8.51$\pm$0.14  &6.5 & 7.04$\pm$0.35 &2.2 & 2.20$\pm$0.01 &1.8 & 1.80$\pm$0.01  \\
\hline
AGN2 &19.0 & 19.89$\pm$0.11 &16.0 & 18.26$\pm$0.77 &2.8 & 2.78$\pm$0.01 &0.8 & 0.82$\pm$0.02  \\
\hline
AGN1 &17.8 & 19.61$\pm$2.20  &  -- & -- &  4.6 & 4.37$\pm$0.06 & 3.1 & 3.27$\pm$0.93  \\
\hline
\end{tabular}
\end{table*}

\begin{figure}
  \includegraphics[width=\columnwidth]{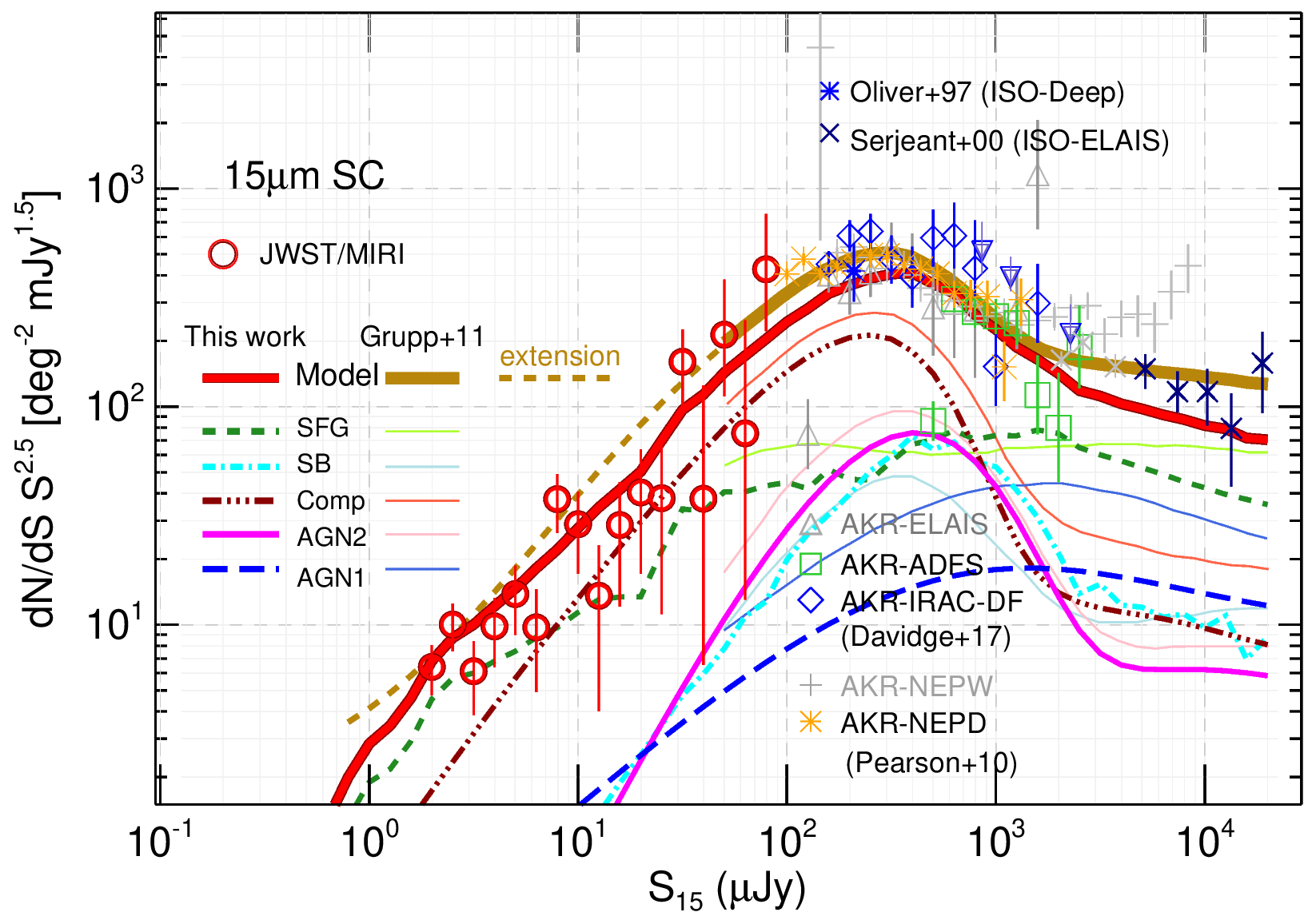}
    \caption{Comparison of 15 $\mu$m SC models between the initial and resultant parameters. Individual populations based on the initial parameters \citep[i.e.,][]{Gruppioni2011} are presented by thin lines and new models with updated parameters are presented by thick lines, as specified in the legend.  The open red circles represent the recent MIR SC (Ling et al. 2022; Wu et al. 2023) obtained from the early release science (ERS) data of the \textit{JWST}.  Green boxes and blue diamonds indicate the source counts on the AKARI Deep Field South (ADF-S), and \textit{Spitzer}/IRAC dark field (IRAC-DF) observed by the AKARI, respectively \citep[]{Davidge2017}.   Crosses (+) and yellow asterisks  indicate the results from the AKARI's NEP-Deep and NEP-Wide fields \citep{Takagi2012, Pearson2010}. The extension is indicated by the dashed line from the light brown curve \citep[model from][]{Gruppioni2011}. This extension is created by summing up each extrapolation for five populations.
      } 
    \label{f8_15umSCcmp}
\end{figure}

\begin{figure*}
  \includegraphics[width=\textwidth]{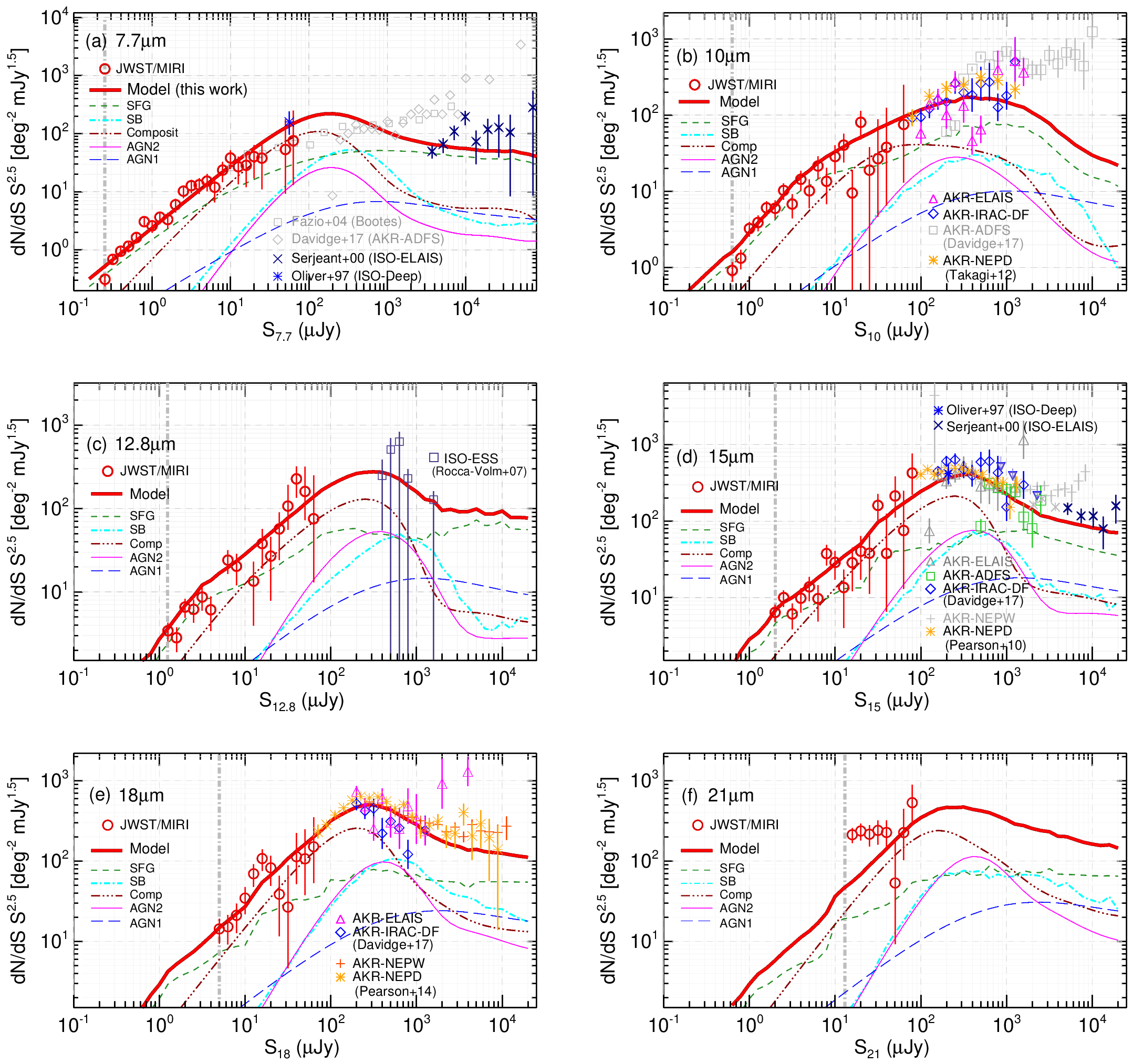}
    \caption{Observed source counts (SCs) from various works and best-fit models at the six bands (F0700W, F1000W, F1280W, F1500W, F1800W, and F2100W) of the \textit{JWST}/MIRI. Open red circles show the recent MIR SC \citep[][Wu et al. 2023]{Ling2022} using early release science (ERS) data of the \textit{JWST}. The vertical dot-dashed grey line indicates 80$\%$ completeness limit. Magenta triangles (in the panels for 10 and 18 $\mu$m SCs), green boxes (in the panels for 15 $\mu$m SCs), and blue diamonds (in the panels for 10, 15, and 18 $\mu$m SCs) show the source counts on the ELAIS field, the AKARI Deep Field South (ADF-S), and \textit{Spitzer}/IRAC dark field (IRAC-DF) observed by the AKARI, respectively \citep[]{Davidge2017}.  ISO data in 7 and 15 $\mu$m are from Oliver et al. (1997) and \citet{Serjeant2000}.  Yellow asterisks (10, 15, and 18 $\mu$m SCs) and red crosses (18 $\mu$m SCs) indicate the results on the AKARI's NEP-Deep and NEP-Wide field \citep{Takagi2012, Pearson2010, Pearson2014}.  The grey colour indicates that they are excluded in the curve-fitting procedure (see Sec. 3.3). The thick red curve in each panel shows the model fitted to all source count results, which is the sum of all the contributions from five different types of galaxies. The green dashed line indicates the SFG, the cyan dot-dashed line indicates the starburst (SB), the dark-red dot-dashed line indicates the composite (LLAGN), the magenta line indicates the AGN type-2, and the blue dashed line indicates the AGN type-1. } 
    \label{f9_6bandsfit}
\end{figure*}

\begin{figure*}
  \includegraphics[width=0.95\textwidth]{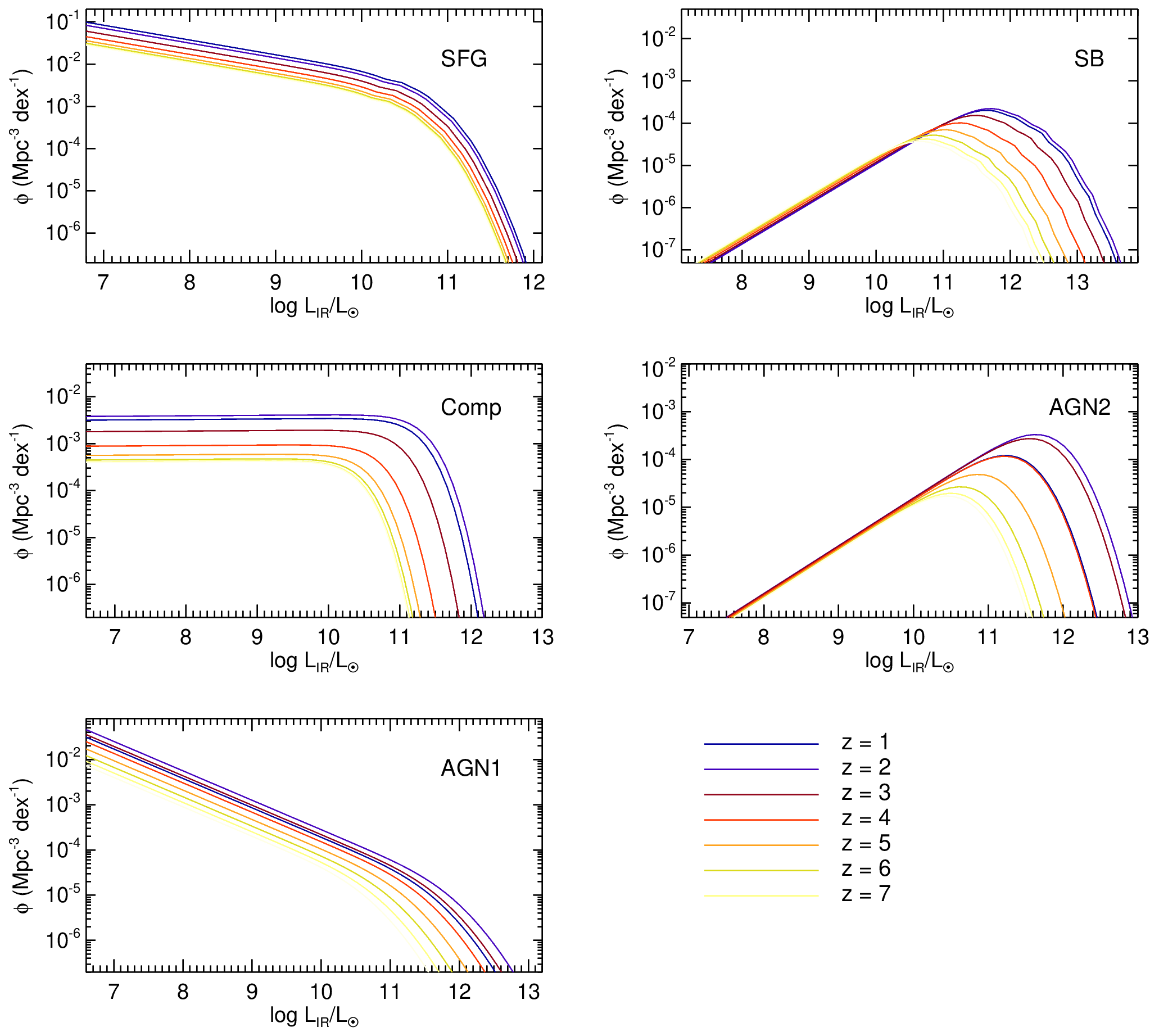}
    \caption{Total IR luminosity functions for five galaxy populations at the redshift $z$=1,2,3, ..., 7, which are obtained based on the best-fit parameters to the observed number counts at six MIRI bands. See sec 4.2.      } 
    \label{f10_LIRs}
\end{figure*}

\subsection{Model fit to the observed source counts  }

We perform a comparison between our model and the observed source counts by utilizing a curve-fitting procedure. Initially, we define functions that generate source count models for each \textit{JWST}/MIRI band, incorporating galaxy spectral energy distributions (SEDs) and initial LLFs that follow the backward evolution of luminosity and density (as described in equations 2 and 3, in Sec 3.2). In total, we prepare six user-defined functions for the six MIRI bands. These source count models, along with the evolution parameters, are then fitted simultaneously to the observed source counts in each band to obtain the best-fit results. To facilitate this process, we employ the `\textsc{MPFIT}' package  \citep[][]{Markwardt2009, Markwardt2012}, which utilizes the Levenberg-Marquardt technique for solving the least-squares problem. 

As an initial guess/input in our model, we adopt the evolution parameters provided by \citet[][]{Gruppioni2011} -- their best estimates for the parameters as listed in Table 2. To prevent potential instabilities and catastrophic runaways of parameters during the fitting, we impose limits on the parameter ranges based on preliminary trials (e.g., testing a single parameter variation with the others fixed at the initial values) using the following -- larger values for the parameters $A_L$, $A_{\Phi}$, and $\kappa$ results in the SC model shifting upward, while increasing $\omega$ causes the overall height of the SC model to decrease. Furthermore, $\omega$ should not be zero, as it appears as the denominator in the error functions in equations (2) and (3). Additionally, it should be also noted that a single population alone, without considering other population components, cannot fit the observed data points, highlighting the importance of multiple components in the model, although these are somewhat arbitrary.

For the data points to fit, we take the observed source counts from the literature mentioned in Sec.2. The errors in the observed source counts are also taken as input, which are used as weight in the curve fitting procedure.  To construct a comprehensive dataset, we combine the \textit{JWST} source counts with those from previous works mentioned in Sec.2.  Specifically, the AKARI 11$\mu$m (S11) source counts are combined with the source count at 10$\mu$m (F1000W of the \textit{JWST}).  The ISO 12$\mu$m source counts \citep{Roc-Volm2007}  are combined with the \textit{JWST} 12.8$\mu$m (F1280W) source count. The previous 15 and 18$\mu$m source counts are merged with the F1500W  and F1800W source counts of the \textit{JWST}, respectively.

\citet[][in Fig. 1 and Fig. 2]{Fazio2004} demonstrated that most of the detected sources in the bright ranges  (8$\mu$m) are stars. \citet[][in Table 2]{Pearson2010} showed that even at the MIR bands (e.g., 15 $\mu$m), the stellar sources occupy more than 50 $\%$ (at 18.6 mJy) of the source detection. 
\citet{Davidge2017}  pointed out that  Fazio et al. (2004) estimated stellar contribution statistically by modelling because the observed area was not fully covered by optical bands at the time. It should be noted that the reliability of the stellar subtraction method can impact the SCs at the brightest flux ranges. Consequently, we did not include these data points in our fitting process, as indicated by grey data points (in Fig. \ref{f9_6bandsfit}).  After the fitting, we see that the data points presented in grey exceed the best-fit model in the brightest flux range.

\begin{figure*}
  \begin{center}
  \includegraphics[width=0.8\textwidth]{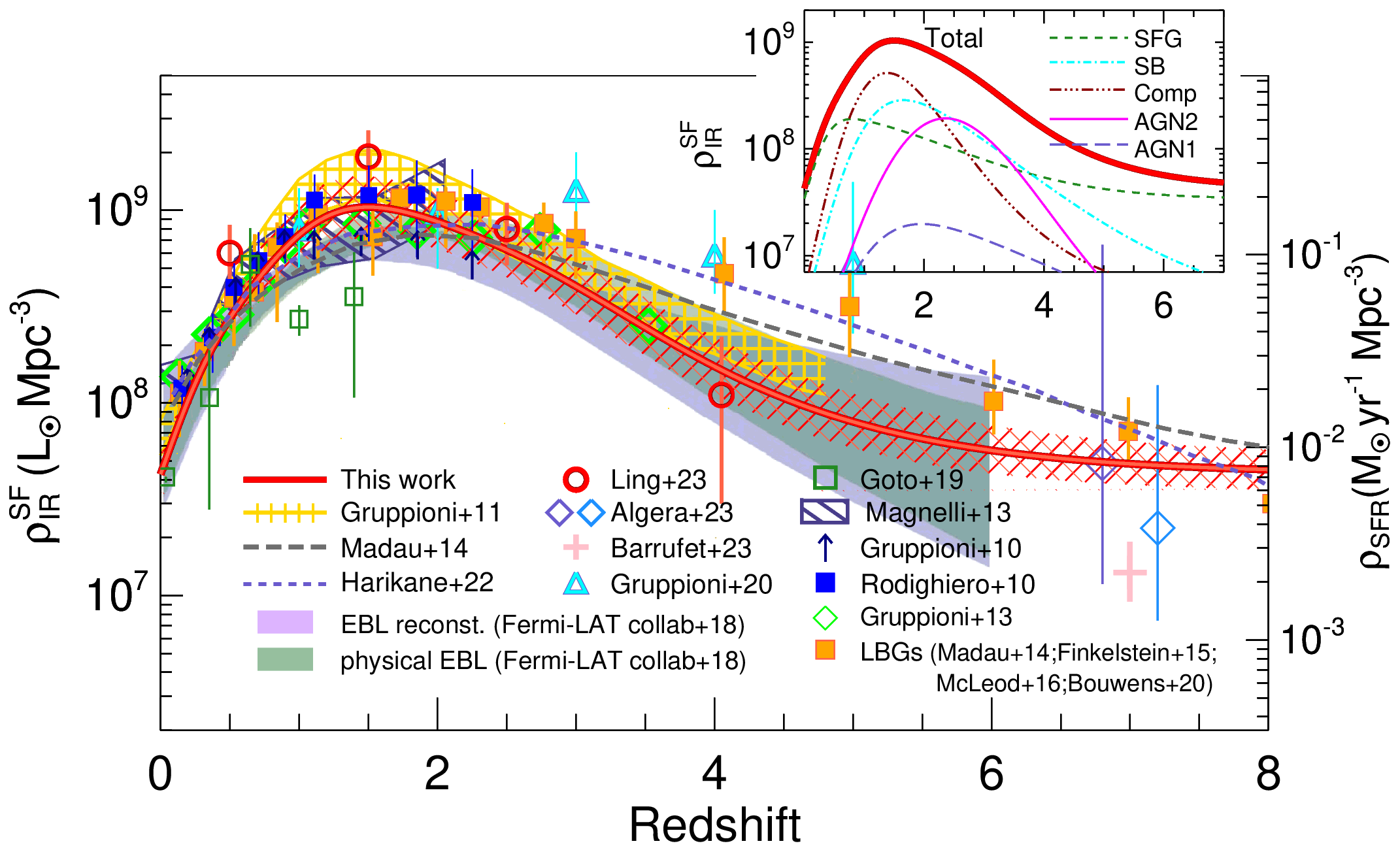}
  \caption{ Redshift evolution of total IR luminosity density (cosmic star-formation history, CSFH) derived from our evolution model (red). The thick red curve represents this work. Other models are compared, as indicated by the yellow meshed area \citep{Gruppioni2011}, grey dashed curve \citep{Madau2014}, blue dotted curve \citep{Harikane2022}, and purple/green shaded regions \citep{Fermi_LAT2018}.  Observed cosmic SF densities are from the literature -- red open circle (Ling et al. submitted, JWST/CEERS), diamonds (Algera et al. 2023, REBELS) , a pink cross (Barrufet el al. 2023),  triangles (Gruppioni et al. 2020, ALPINE), green diamonds (Gruppioni et al. 2013), upward arrows \citep{Gruppioni2010}, navy region (Magnelli et al. 2013), and blue filled squares \citep{Rodighiero2010}. The orange boxes are   taken from \citet{Madau2014},Finkelstein et al. (2015), McLeod et al. (2016), and Bouwens et al. (2020).   The red meshed area indicates our error range of the best fit. A small panel on the top-right shows the contribution of five different galaxy types. All results are converted based on the Salpeter (1995) IMF. See sec 4.2.
   } 
    \label{f11_CSFH}
   \end{center}    
\end{figure*}

\begin{figure}
 \includegraphics[width=\columnwidth]{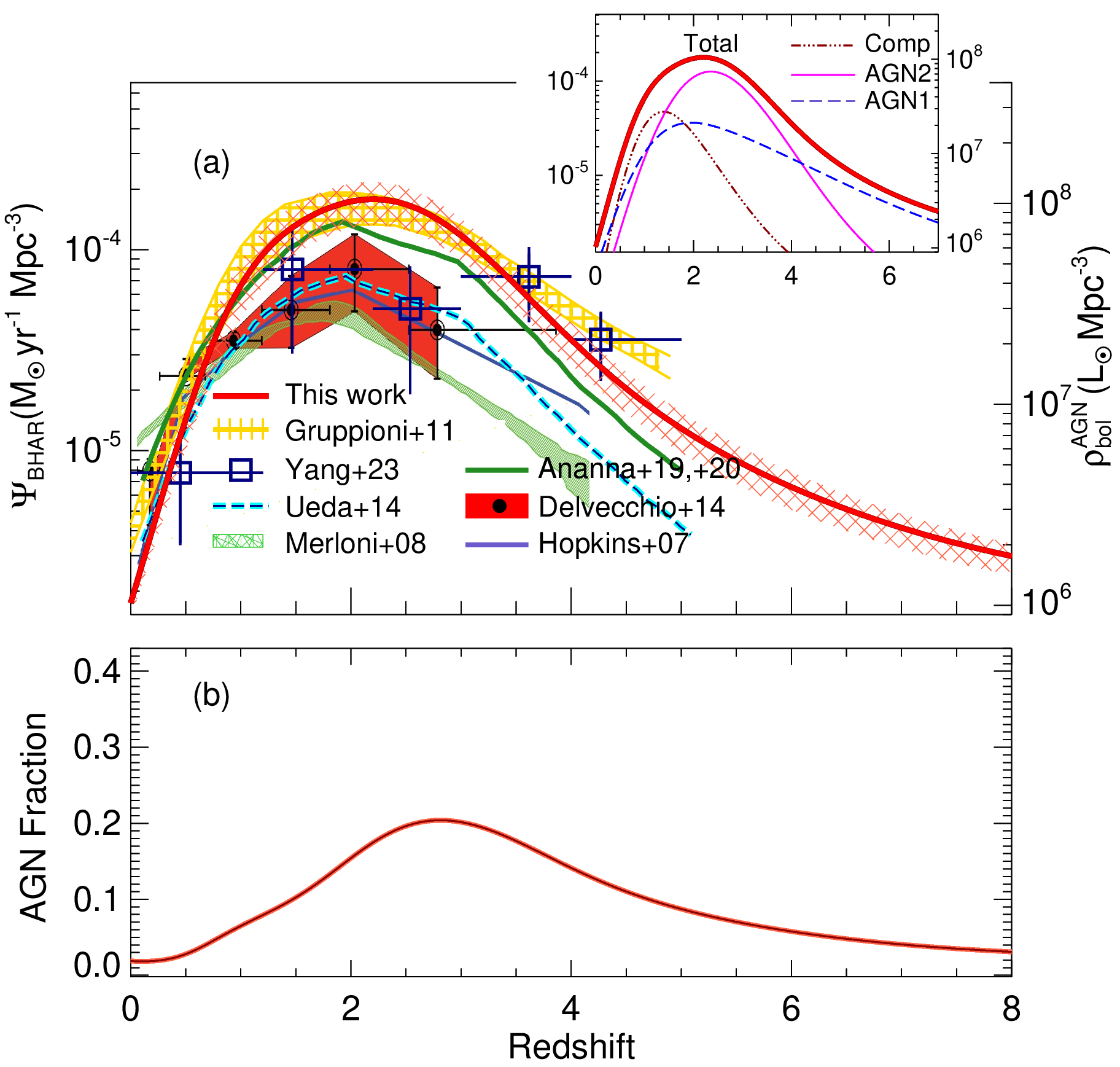}
 \caption{ (a) Redshift evolution of the black hole accretion rate (BHAR). Composite, AGN-2, and AGN-1 populations are used whereas SFG and SB are not used to exclude pure SF activity. The red curve indicates our estimation. The red meshed area represents the fitting error reflected on this plane. A small panel on the top-right shows the contribution from three populations. \citet[][yellow region]{Gruppioni2011} is compared. Open squares indicate the measurements from Yang et al. (2023). The green solid line indicates Ananna et al. (2019, 2020). The Cyan dashed line indicates Ueda et al. (2014) The filled red area with black data point, green hatched area, and blue solid line indicate BHAHs presented by \citet{Delvecchio2014}, \citet{Merloni2008}, and \citet{Hopkins2007}, respectively.     (b) The overall fraction of the cosmic AGN activity, the ratio of $\rho_{\rm 1-1000}^{\rm AGN}$ with respect to the total energy density ($\rho_{\rm 1-1000}^{\rm SF} + \rho_{\rm 1-1000}^{\rm AGN}$). 
  }
 \label{f12_BHAH}
\end{figure}

\section{Results and Discussion}
\label{sec4}

\subsection{Best-fit model and new evolution parameters}

We perform a simultaneous fitting of the source counts across all six MIRI bands, enabling us to derive new evolution parameters (see $A_L$, $A_{\Phi}$, $\kappa$, and $\omega$ in Table \ref{table1}) and obtain the best-fit models for the source count in each band. In Fig. \ref{f5_evLLFs2},  we show new evolution models for $15\mu$m LLFs and compare them with the initial LLFs to illustrate how the new parameters have changed the initial evolution model. The solid curves represent new LFs, and dotted curves represent the initial LFs, at redshifts $z=1, 2, 3, ..., 7$, and 8 (see colour bar on the bottom right). Additionally, to aid understanding, we add insets to show how luminosity (i.e., equation 2, solid line) and density (i.e., equation 3, dot-dashed) evolutions are changed explicitly -- the evolution with initial parameters is given in black, and new evolution models are shown in magenta.  In particular, in Fig. \ref{f6_LF15obs}, we showcase a comparison of our model LFs (at different redshifts), against the observed LFs, which are measurements from the AKARI's NEP-Wide field \citep{Kim2012}, Spitzer observation on the CDFS field (Le Floc'h et al. 2005) and \textit{JWST} CEERS field (Ling et al. submitted). Our model and observed LFs are consistent except for high-z ($z=3 - 5)$ where they show slight deviations. %

Similarly to Fig. \ref{f5_evLLFs2} for LLFs, we present Fig. \ref{f7_evNCs2a} to illustrate how the number counts have changed with the new parameters. The dotted curves represent the initial number counts  
while the solid curves represent the new ones based on the new parameters (presented in Table \ref{table1}). The new parameters exhibit slight deviations from the original values. 
These minor changes result in slightly different models for each galaxy population.  Although these differences may not lead to a significant alteration in the source count model, they enable us to have a slightly different description of the \textit{JWST} source counts.

In Fig. \ref{f8_15umSCcmp}, we present a detailed comparison of our source counts models (red curve) against the model from \citet[]{Gruppioni2011}. The SC model from \citet[][Fig. 5]{Gruppioni2011} is presented in Fig. \ref{f8_15umSCcmp}, using a thick light brown curve.
Overall, the shapes of both models fit well to the characteristic bump near a few 0.1 mJy and show a good agreement, although they show a difference at $S_{\nu}$ > 2mJy.  
For comparison, the slope of the light brown curve \citep{Gruppioni2011} is extended towards the faint end. This extension (indicated by a dashed brown line) is formed by summing up each extrapolation for five populations.  It is slightly higher than our model.  Between 2-- 5$\mu$Jy, both models are consistent, but below $\mu$Jy, extension of \citet{Gruppioni2011} goes higher again.
The updated parameters do not cause a dramatic variation in the final SC model, but they introduce slight variations in each component/population. The combination of these slight differences eventually enables a slightly better description of the \textit{JWST} source counts. The slight change means that the previous parameters have also been remarkably good. Hence, the parameter fitting with the \textit{JWST} data serves as a fine-tuning procedure for the evolution parameters.

In Fig. \ref{f9_6bandsfit}, we show the results of the simultaneous fits to all observed source counts in six MIRI bands.
The resultant best-fit models are given by thick red curves, which are the integrated results of five galaxy populations (based on the best-fit parameters). The distribution of the observed source counts from previous works mainly describes the shape of the bump (see especially 15 and 18 $\mu$m), while the \textit{JWST} SCs show how the behaviour of the faint-end slope has to be fitted. 
Overall, our source counts models peak around 0.1–0.5 mJy. Towards the faint end, they monotonically decrease describing the source counts in the \textit{JWST} bands. 
From the bump ($\sim 0.3$ mJy) towards the bright end ($\sim 20$ mJy), our models slowly decrease or become flat (e.g., 7.7 and 12.8 $\mu$m). We, however, noticed some of the observed source counts (for example, the square and plus symbols in 7.7, 10 and 15$\mu$m panels, presented in grey) rather increase. We regard this related to stellar subtraction. This is the reason we ruled out the SC data from \citet[][Spitzer 8$\mu$m SC]{Fazio2004} for 7.7$\mu$m, ADF-S field data \citep[][]{Davidge2017} for the 7.7 and 10 $\mu$m fitting, and ones from NEPW field \citep{Pearson2005} for 15 $\mu$m.
In panel (d), the triangles are excluded because their completeness is lower than $80\% $. The previous 24$\mu$m source counts \citep[e.g.,][]{Papovich2004} might be available, but we do not use them in the 21$\mu$m fitting procedure because the effective wavelength is quite different.  

Our model shows that the bump shapes are strongly dependent on the composite population, shown by the brown dot-dot-dashed curve in Fig. \ref{f9_6bandsfit}. 
From the bump towards the faint end where the JWST data points are distributed, the shape of the best-fit curve is shaped mostly by the composite and SFG populations. 
Near the faint end ($80\%$ completeness limit is indicated by vertical dot-dashed grey line), SFGs dominate by a few factors on average (the SFG has a dominant influence on the faint and bright end).  
The other populations (i.e., SB, AGN2, and AGN1) do not significantly influence the simultaneous variation of the bump shape and faint end slope (compared to the composite). This is because the relative ratios originated from their different LFs as well as the different evolution styles.   
A larger number of data points around the bump feature in 15$\mu$m (or 18$\mu$m) source counts may exert stronger constraints in the fitting procedure. A possible factor that slightly restricts the flexibility of the fitting process is the imposed constraints on the parameter ranges to prevent catastrophic parameter runaway during the fitting. 
There is scatter in the \textit{JWST} SCs. We regard that one of the reasons for this is the small number of sources in the bins as well as the faint sources near the detection limit (80\% completeness limit). This may be attributed, in part, to cosmic variance. 
Drawing a conclusive remark on the 21$\mu$m source count is not straightforward owing to the lack of data points in the brighter flux range, compared to 15 and 18 $\mu$m counts.  An important point of this work is not the difference from previous work,  but we used detected $\mu$Jy sources — just an extrapolation in previous work — to estimate the LFs and CSFH.

\subsection{CSFH and BHAH}
 
In order to estimate the cosmic evolution of the SF density, we derive the total IR luminosity and IR luminosity functions (IRLFs).
Total IR luminosity  ($L^{\rm tot} _{8-1000}$) is derived by integrating galaxy SED from 8$\mu$m to 1000$\mu$m, based on \citet{Polletta2007}. This contains both SF and AGN contributions. To estimate the SF contribution ($L^{\rm SF}_{8-1000} = \rm{frac} \times  L^{\rm tot} _{8-1000}$), we used the fraction (percentage) provided  by \citet[][Table 3]{Gruppioni2011}. 
The SED of SFG and SB comes from purely SF activities. For the composite, AGN2 and AGN1,  SF contributions are 96\%, 64\%, and 46\%,   respectively. The other parts originate from the AGN activity. Total IR LFs are obtained with the updated evolution parameters, following equations (2) and (3). In Fig. \ref{f10_LIRs}, we show the total IR LFs of the five populations of galaxies.  Based on these new IR LFs, we obtain  comoving IR luminosity density, $\rho_{\rm IR}^{\rm SF}(z)$,  obtained by multiplying luminosity ($L_{8-1000}^{\rm SF}$) by density $\phi$($L_{8-1000}$), described as,  
\begin{equation}
\label{eqSFD}
 \rho_{\rm IR}^{\rm SF}(z)=\int_{0}^{\infty} L_{8-1000}^{\rm SF} \phi(L_{8-1000}) \, {\rm d}\, {\rm log}L_{8-1000}
\end{equation} 
excluding the AGN contribution to the total IR energy. 
This can be directly converted to the star formation rate (SFR) density as a function of $z$, using the conversion of \citet{Kennicutt1998}, $\rho_{\rm SFR} = 1.7 \times 10^{-10} \, \rho_{\rm IR}^{\rm SF} \, ({\rm M}_{\odot} {\rm yr}^{-1} {\rm Mpc}^{-3}$)  based on the Salpeter initial mass function (Salpeter 1955). This is a crucial tool to understand galaxy evolution in terms of cosmic star formation history (CSFH) \citep{Goto2010, Goto2019}.

We derive the IR luminosity density as a function of redshift,  as estimated from our model and compare it with those from other works.  
In Fig. \ref{f11_CSFH}, We show a comparison with a model from \citet[][yellow meshed area]{Gruppioni2011}. The grey dashed curve indicates the cosmic star formation history from \citet{Madau2014}. The Blue dotted line indicates the model curve from \citet{Harikane2022}, which compares with their high-$z$ ($z\sim13$) galaxy candidates. Purple and green shaded areas show the empirical reconstruction of extragalactic background light (EBL)  and physical EBL model, respectively \citep{Fermi_LAT2018}.

We also compare with observed cosmic SFR densities from the literature.  The red open circles represent the observed SFR density from Ling et al. (2023, based on the CEERS field IRLFs).  Open diamonds indicate the observational constraints from the ALMA (Algera et al. 2023, REBELS). The pink cross indicates the measurement from the HST-dark, \textit{JWST}/NIRcam galaxies (Barrufet et al. 2023). 
Open triangles are from Gruppioni et al. (2020, ALPINE-ALMA survey). Navy dashed region is from  Magnelli et al. (2013, Herschel-PACS survey on GOODS field). Upward arrows are from \citet{Gruppioni2010}, which used dusty galaxies detected by Herschel ($z\sim$3). Filled blue boxes indicate \citet[]{Rodighiero2010}, based on a combination of \textit{Spitzer} surveys of VVDS-SWIRE (de la Torre et al. 2007) and
GOODS (Giavalisco et al. 2004) fields. Orange-filled boxes are taken from \citet{Madau2014}, Finkelstein et al. (2015), McLeod et al. (2016), Bouwens et al. (2020), which indicate the constraints from Lyman break galaxies (LBGs). 
Magnelli et al. (2013) and Grup-
pioni et al. (2020) used Chabrier (2003) IMF, while  \citet{Rodighiero2010} and \citet{Gruppioni2011} used Salpeter (1955) IMF.   All results are converted based on the Salpeter (1955) IMF. 

Our model (red curve) shows a rapid increase up to $z\sim$1, followed by a peak at $1<z<2$. It decreases sluggishly from $z\sim2$ toward higher-$z$, becoming flat at $z>6$. From the current epoch to around the $z\sim$1, all works show a good agreement.  While the pink cross  (Barrufet et al. 2023) at $z=5$ and the blue dotted line  \citep[][]{Harikane2022} around $z=4.5$ show the largest differences  (about factor three), our model is consistent with the other recent observational constraints in error ranges.

The accretion rate of supermassive black holes (SMBHs) can be estimated using  $\Phi(L^{\rm AGN}_{\rm bol}, z)$, where $L^{\rm AGN}_{\rm bol}$ ($ = \epsilon_{\rm rad} \dot{M}c^2$) is the intrinsic bolometric luminosity, produced by SMBH accreting at a rate of $\dot{M}$ with a radiative efficiency $\epsilon_{\rm rad}$.  For $\epsilon_{\rm rad}$, we take the same value as \citet{Gruppioni2011} and use 0.1 \citep{Hopkins2007}.   
Considering the majority is emitted at long wavelengths, we approximate the bolometric luminosity ($L^{\rm AGN}_{\rm bol}$) to the IR luminosity integrated between 1 and 1000 $\mu$m ($L^{\rm AGN}_{\rm 1-1000}$), following \citet{Gruppioni2011}. 
For IR luminosity integrated in this wavelength range ($L^{\rm tot}_{\rm 1-1000}$), the AGN contributions are 9\%, 44\%, and 69\% for the composites, AGN2, and AGN1 types, respectively, as given by \citet[][Fig. 10 and Table 3]{Gruppioni2011}. 
Here, we estimate AGN luminosity ($L^{\rm AGN}_{\rm 1-1000}$) based on its contribution (fraction) to $L^{\rm tot}_{\rm 1-1000}$. 
SFGs and starburst are purely star-forming populations, therefore, AGN fractions are 0. The black hole accretion rate (BHAR) is described as 
\begin{equation}
\label{eqBHAR}
 \Psi_{\rm BHAR}(z)=\int_{0}^{\infty} \frac{(1-\epsilon_{\rm rad})(\rm BC) L_{1-1000}^{\rm AGN}} {\epsilon_{\rm rad}c^2} \phi(L_{1-1000}) {\rm d}\, {\rm log}L_{1-1000},
\end{equation}
where BC is the bolometric correction to the 1-- 1000 $\mu$m IR luminosity depending on the SED type, and $L^{\rm AGN}_{1-1000}$ indicates the 1--1000 $\mu$m IR luminosity due to the AGN.   For the bolometric correction (BC), we take the same values as \citet{Gruppioni2011}, and use BC $\sim$ 1.5 for the AGN1 and AGN2, and BC $\sim$ 2 for the composite (LLAGN). These bolometric corrections are first-order empirical estimates derived in Pozzi et al. (2007), where the broad-band SEDs of a sample of X-ray-selected AGNs have been studied.

Fig. \ref{f12_BHAH} shows the black hole accretion rate derived from our model and comparison with other results. 
In \citet{Merloni2008}, the hard X-ray luminosity function is taken as a tracer of the AGN growth rate distribution. They estimated black hole accretion density and presented a synthesis model for AGN evolution based on the hard X-ray LF.    \citet{Hopkins2007} derived the quasar bolometric LF from X-ray data.  From $z=0$ to $z=1$, all works are consistent and roughly agree. However, at $z>1.5$ differences appear. 
Compared to other works, our model and \citet{Gruppioni2011} are higher by a few factors. This might be because of the different methods using different data sets. 

The physical meaning of AGN energy density is the IR emission originating from the circum-nuclear dusty material that intervenes a fraction of the optical/UV radiation from the centre.  
In this wavelength range (1-1000$\mu$m) along the SED, to see the AGN contribution against the SF contribution (or total amount), we calculated energy density
$\rho_{1-1000} ^{\rm AGN}(z)=\int L_{1-1000}^{\rm AGN} \phi(L_{1-1000})~{\rm d} {\rm log}L_{1-1000}$ and $\rho_{1-1000}^{\rm SF}(z)=\int L_{1-1000}^{\rm SF} \phi(L_{1-1000})~{\rm d} {\rm log}L_{1-1000}$ using the percentages in \citet[][Table 3]{Gruppioni2011}, and compared them. 
Panel (b) shows the fraction of AGN contribution, which is the ratio of $\rho_{1-1000}^{\rm AGN}$ with respect to the total energy density ($\rho_{1-1000}^{\rm SF} + \rho_{1-1000}^{\rm AGN}$). While SF activity peaks at $1<z<2$ (Fig. \ref{f11_CSFH}), the black hole activity (Fig. \ref{f12_BHAH}a) peaks at $2<z<2.5$. However, the actual role of AGN is most significant at higher-$z$ ($2.5<z<3$) based on  Fig. \ref{f12_BHAH}b.

Our estimates incorporate MIR sources that are several tens of times fainter, thanks to the \textit{JWST}, which were only predicted or extrapolated in previous studies. As the slight parameter adjustments show, this work demonstrates the remarkable accuracy of predictions and estimations made in the previous studies.

\section{Summary}
\label{sec5}

To fully understand galaxy evolution, it is ideal to have redshift information and investigate galaxy luminosity functions. However, at the time of writing, we do not have photo-z measurements for all the JWST data we used in this work. Some fields lack enough optical data needed for accurate photo-z. Therefore, we took a model-based approach to interpret galaxy number counts and extract information on galaxy evolution.
This approach allows us to obtain a physical interpretation of the MIR source counts (Ling et al. 2022, Wu et al. 2023) in a wide flux range using extremely faint galaxies, and constrain cosmic star-formation history and black hole accretion history.
Following \citet[][]{Gruppioni2011}, we used the backward evolution of the parameterized LLFs for five representative galaxy populations at the \textit{JWST}/MIRI bands.
We fit our model to the source counts from the \textit{JWST}  as well as previous works from ISO, AKARI, and Spitzer \citep[Oliver et al. 1997;][]{ Serjeant2000, Roc-Volm2007, Pearson2010, Pearson2014, Davidge2017}.
By simultaneously fitting the source counts at six mid-infrared bands, we obtained the best-fit evolutions of MIR LFs for each of the five types of galaxies.
These parameters gave us inferred CSFH and BHAH.
From the current epoch to $1<z<2$, the obtained CSFH shows good agreement with previous works,  at higher-$z$, however, uncertainties become large and the overall decreasing trends are not the same, yet consistent within error ranges.
BHAH estimates are consistent with previous estimates in terms of peak density; however, our estimate is slightly higher compared to others at $z>1$.
These differences seem to be due to the different data and methods used.
Our results take advantage of IR data that is less affected by dust extinction.
Thanks to the \textit{JWST}, this work is based on infrared galaxies that are several tens of times fainter than those detectable by previous IR telescopes, shedding light on the accuracy of previous studies.

\section*{Acknowledgements}

The authors thank the anonymous referee for many constructive comments, which improved the paper much.
TG acknowledge the support of the National Science and Technology Council of Taiwan through grants 108-2628-M-007-004-MY3, 110-2112-M-005-013-MY3, 111-2112-M-007-021, 111-2123-M-001-008-, 112-2112-M-007-013, and 112-2123-M-001-004-. 
TH acknowledges the support of the National Science and Technology Council of Taiwan through grants 110-2112-M-005-013-MY3, 110-2112-M-007-034-, and 111-2123-M-001-008-, and 112-2123-M-001-004-.
SH acknowledges the support of The Australian Research Council Centre of Excellence for Gravitational Wave Discovery (OzGrav) and the Australian Research Council Centre of Excellence for All Sky Astrophysics in 3 Dimensions (ASTRO 3D), through project number CE17010000 and CE170100013, respectively.
This work is based on observations made with the NASA/ESA/CSA James Webb Space Telescope. The data were obtained from the Mikulski Archive for Space Telescopes at the Space Telescope Science Institute, which is operated by the Association of Universities for Research in Astronomy, Inc., under NASA contract NAS 5-03127 for \textit{JWST}. These observations are associated with the program ERO.


\section*{Data Availability}
Early Release Science (ERS) data obtained by the James Webb telescope, \textit{JWST} MIRI are publicly available at \url{https://www.stsci.edu/jwst/science-execution/approved-programs/webb-first-image-observations}.
Other data described in this article will be shared upon reasonable request to the corresponding author.
 


\bibliographystyle{mnras}

\bibliography{SC_model_JW} 








\bsp	
\label{lastpage}
\end{document}